\documentclass[preprint,12pt]{elsarticle}
\usepackage{latexsym, graphicx, epsfig, amsmath, amssymb, amsfonts}
\usepackage{amsmath}
\usepackage{amssymb}
\usepackage{graphicx}
\usepackage{subfig}
\usepackage{tabu}
\usepackage{adjustbox}
\usepackage{bbold}
\usepackage{bm}
\usepackage{textcomp}
\usepackage[thinc]{esdiff}
\usepackage[dvipsnames]{xcolor}
\usepackage{comment}
\usepackage{textgreek}
\usepackage{caption}
\usepackage{lineno}
\usepackage{multirow}
\usepackage{adjustbox}

\usepackage{siunitx}
\usepackage{xcolor}
\usepackage{booktabs}
\usepackage{caption}
\usepackage{float}
\usepackage{color,soul}

\journal{the Journal}
\begin{document}
\begin{frontmatter}

\title{Mechanical Metamaterials Fabricated from Self-assembly:               
A Perspective
}

\author{Hanxun Jin\textsuperscript{a}\corref{cor1}}
\author{Horacio D. Espinosa\textsuperscript{b}\corref{cor2}}
\cortext[cor1]{Corresponding authors. E-mail:  hjin@caltech.edu (H.J.)}
\cortext[cor2]{Corresponding authors. E-mail:  espinosa@northwestern.edu (H.D.E.)}

\address[1]{Division of Engineering and Applied Science, California Institute of Technology, Pasadena, CA 91125}
\address[2]{Department of Mechanical Engineering, Northwestern University, Evanston, IL 60208}

\begin{abstract}

Mechanical metamaterials, whose unique mechanical properties stem from their structural design rather than material constituents, are gaining popularity in engineering applications. In particular, recent advances in self-assembly techniques offer the potential to fabricate load-bearing mechanical metamaterials with unparalleled feature size control and scalability compared to those produced by additive manufacturing (AM). Yet, the field is still in its early stages. In this perspective, we first provide an overview of the state-of-the-art self-assembly techniques, with a focus on the copolymer and colloid crystal self-assembly processes. We then discuss current challenges and future opportunities in this research area, focusing on novel fabrication approaches, the need for high-throughput characterization methods, and the integration of Machine Learning (ML) and lab automation for inverse design. Given recent progress in all these areas, we foresee mechanical metamaterials fabricated from self-assembly techniques impacting a variety of applications relying on lightweight, strong, and tough materials.
\end{abstract}


\end{frontmatter}

\newpage
\section{Introduction}

The concept of Materials by Design has recently emerged as an innovative approach to fabricating multi-functional metamaterials with unprecedented properties through the strategic design of their material constituents and architecture at various scales \cite{xia2022responsive,montemayor2015materials}. Mechanical metamaterials, alternatively termed architected metamaterials, exhibit distinct mechanical properties governed by their architectural features rather than their material constituents \cite{bauer2017nanolattices, surjadi2019mechanical}. For example, architected materials made from polymer, metals, or ceramics with nano- and micro-scale features, exbibit enhanced strength and stiffness \cite{zhang2019lightweight,jin2023ultrastrong}, superior mechanical recoverability \cite{meza2014strong, lin2020folding}, and impact resilience \cite{portela2021supersonic}. The advancement of metamaterials can be attributed in part to the development of rapid fabrication tools like AM \cite{vyatskikh2018additive}, which can fabricate complex 3D architectures through a layer-by-layer approach. Common AM techniques that can be used to fabricate mechanical metamaterials include direct ink writing (DIW), vat-photopolymerization techniques like digital light processing (DLP) and two-photon polymerization (TPP), and powder bed fusion methods \cite{gibson2021additive, frazier2014metal,chen2022microstructural, wong2012review}. 

Until now, the fabrication of mechanical metamaterials has predominantly relied on AM techniques. While these techniques offer rapid production and custom design, they inherently limit the structural resolution. For example, TPP can fabricate structures with resolutions near 100 nm, resulting in unit cell sizes at the microscale \cite{o2023two}. Consequently, using AM to fabricate mechanical metamaterials with sub-micron scale unit cell dimensions remains challenging. Moreover, scaling nano- and micro-scale features to sizes relevant to engineering applications (ranging from millimeters to meters) demands substantial fabrication time \cite{carlotti2019functional}. Overcoming these challenges would create invaluable opportunities for nano-architected metamaterials with unparalleled resolution, ushering in unprecedented mechanical properties. Recent advances in nanotechnology have introduced an alternative fabrication method for metamaterials, harnessing the power of nanoscale self-assembly processes \cite{mai2012self, grzelczak2010directed}. While self-assembly has been widely employed to fabricate photonic crystals \cite{cai2021colloidal, vignolini20123d} and plasmonic metamaterials \cite{wang2016programming}, its application in the fabrication of load-bearing mechanical metamaterial is still nascent. In this perspective paper, we first provide a concise review of the fabrication and characterization methods of self-assembled mechanical metamaterials, focusing on copolymer and colloidal crystal self-assembly techniques. Subsequently, we outline our perspective for future opportunities and forthcoming challenges in this field.

\section{A brief review on self-assembly methods}

Self-assembly is a process where individual components autonomously organize into ordered patterns or structures driven by local interactions \cite{whitesides2002self}. These interactions can be based on different principles through multiple scales, like chemical potentials and physical forces. One advantage of self-assembly compared to AM is scalability, which allows for synthesizing materials with homogeneous mechanical properties over large dimensions. In this section, we first briefly review copolymer self-assembly techniques and then delve deeper into the latest advancements in colloidal crystal self-assembly methods.

\subsection{Copolymer self-assembly}

\begin{figure}[!ht]
    \centering
    \includegraphics[width=1.0\textwidth]{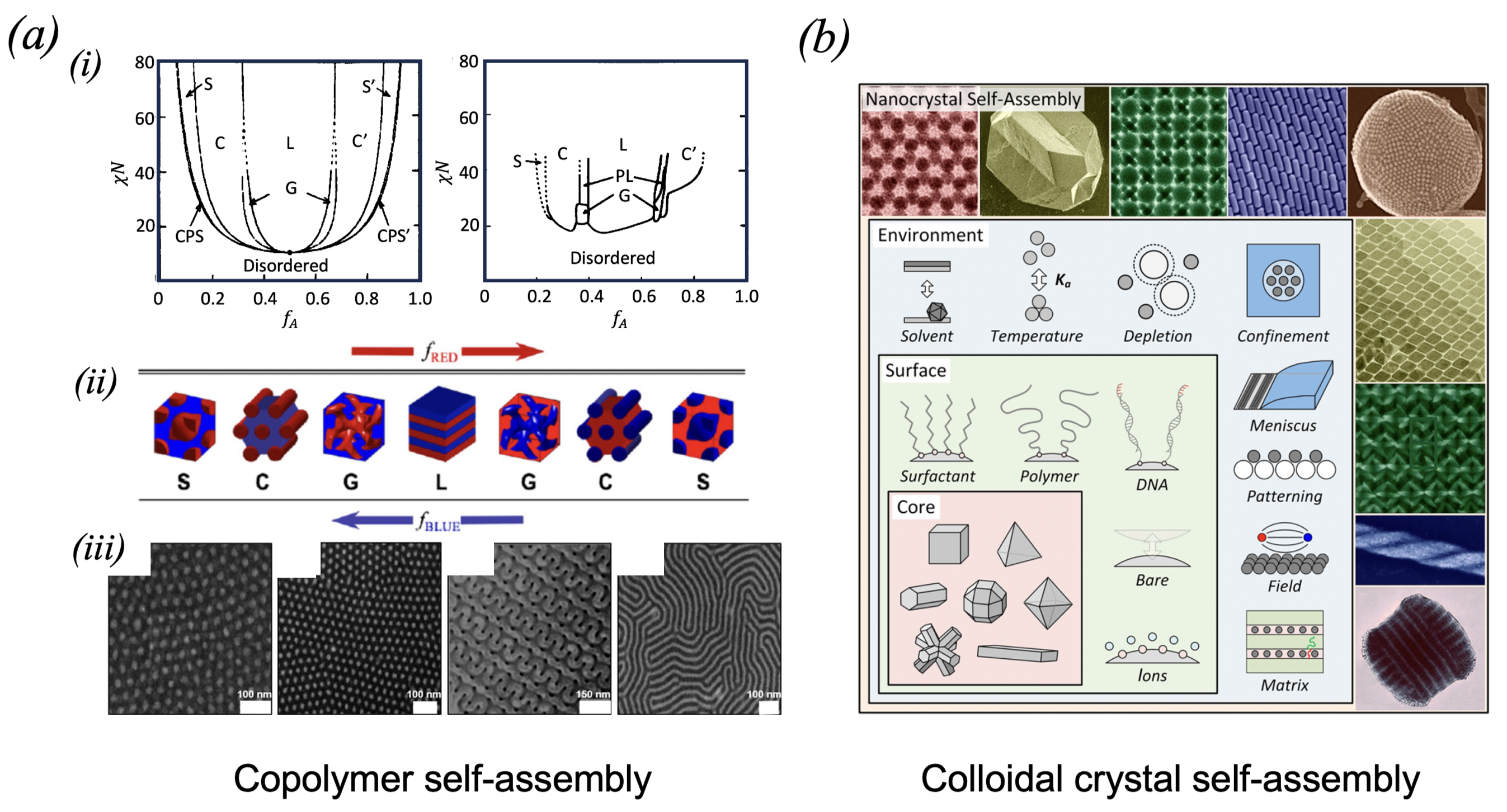}
    \caption{Mechanical metamaterials fabricated from two self-assembly techniques. (a) Block copolymer self-assembly: (i) block copolymer phase diagram calculated from self-consistent mean-field theory and experiments (Reprinted with permission from \cite{bates1999block}. Copyright 1999, American Institute of Physics). (ii) The equilibrium self-assembled morphologies, Spherical (S), Cylindrical (C), Gyroid (G), and Lamellar (L). (iii) SEM images of different block copolymer phases (Reprinted with permission from \cite{cummins2020enabling}. Copyright 2020, Elsevier). (b) Colloidal nanocrystal self-assembly: In this technique, colloidal crystals are functionalized with a layer of surface ligands (e.g., polymer, DNA, surfactant) and then self-assembled into nano- and micro-structures, which are controlled by interparticle interactions, geometric constraints, and environment (Reprinted with permission from \cite{boles2016self}. Copyright 2016, ACS).}
    \label{fig:abstract}
\end{figure}

A copolymer is a type of polymer synthesized from two or more distinct monomer species \cite{young2011introduction}. In block copolymers, the thermodynamically incompatible monomers are covalently linked, which produces unique micro- and nano-scale phase-separated morphologies \cite{bates1990block, bates1999block}. Such interesting morphologies endow copolymers with unique mechanical properties such as exceptional dynamic toughness \cite{jin2022dynamic,jin2022understanding,jin2022big, kim2021dynamic}. The phase separation morphology is primarily governed by the Flory-Huggins parameter \cite{flory1942thermodynamics, huggins1941solutions}, $\chi$, the degree of polymerization, $N$, and the volume fraction of each component $i$, $f_i$. As illustrated in \textbf{Fig.1(a)}, by tuning these parameters and other specific solvent environments or external fields, the mixture can self-assemble into a variety of ordered structures like spheres, cylinders, lamellae, and disordered gyroid phases. These controllable self-assembled polymeric nanoarchitectures can be employed as nanolithography templates, facilitating the production of advanced nano- and micro-scale architectures \cite{cummins2020enabling, hu2014directed, park2003enabling,vukovic2013block}. For example, using copolymer self-assembly followed by electro-deposition of nickel, double gyroid films with a unit cell of $\sim$ 45 nm and feature size $\sim$ 10 nm were fabricated \cite{khaderi2017indentation}. Notably, nanoindentation tests revealed that the strength of this double gyroid nickel structure approaches the theoretical strength of bulk nickel \cite{khaderi2017indentation}, attributed to the absence of dislocations, a well-known size scale feature observed in metallic nanowires \cite{filleter2012nucleation, bernal2015intrinsic}. Beyond copolymer templates, other self-assembly processes, like polymeric emulsions, can produce porous bi-continuous templates \cite{silverstein2014emulsion}. For example, a nano-labyrinthine shell-based material was fabricated through the spinodal decomposition of an epoxy resin emulsion \cite{portela2020extreme}. This epoxy spinodal structure was coated with Al2O3 through atomic layer deposition. Afterward, the epoxy was removed using Oxygen plasma ashing, leading to a unique centimeter-scale spinodal architecture. Through nanomechanical experiments and finite element simulations, a strong mechanical resilience was revealed stemming from its nodeless architecture and low principal curvature, which effectively minimizes local stress concentration.

\subsection{Colloidal nanocrystal self-assembly: from solid to architected superlattices}

\begin{figure}[!ht]
    \centering
    \includegraphics[width=1.0\textwidth]{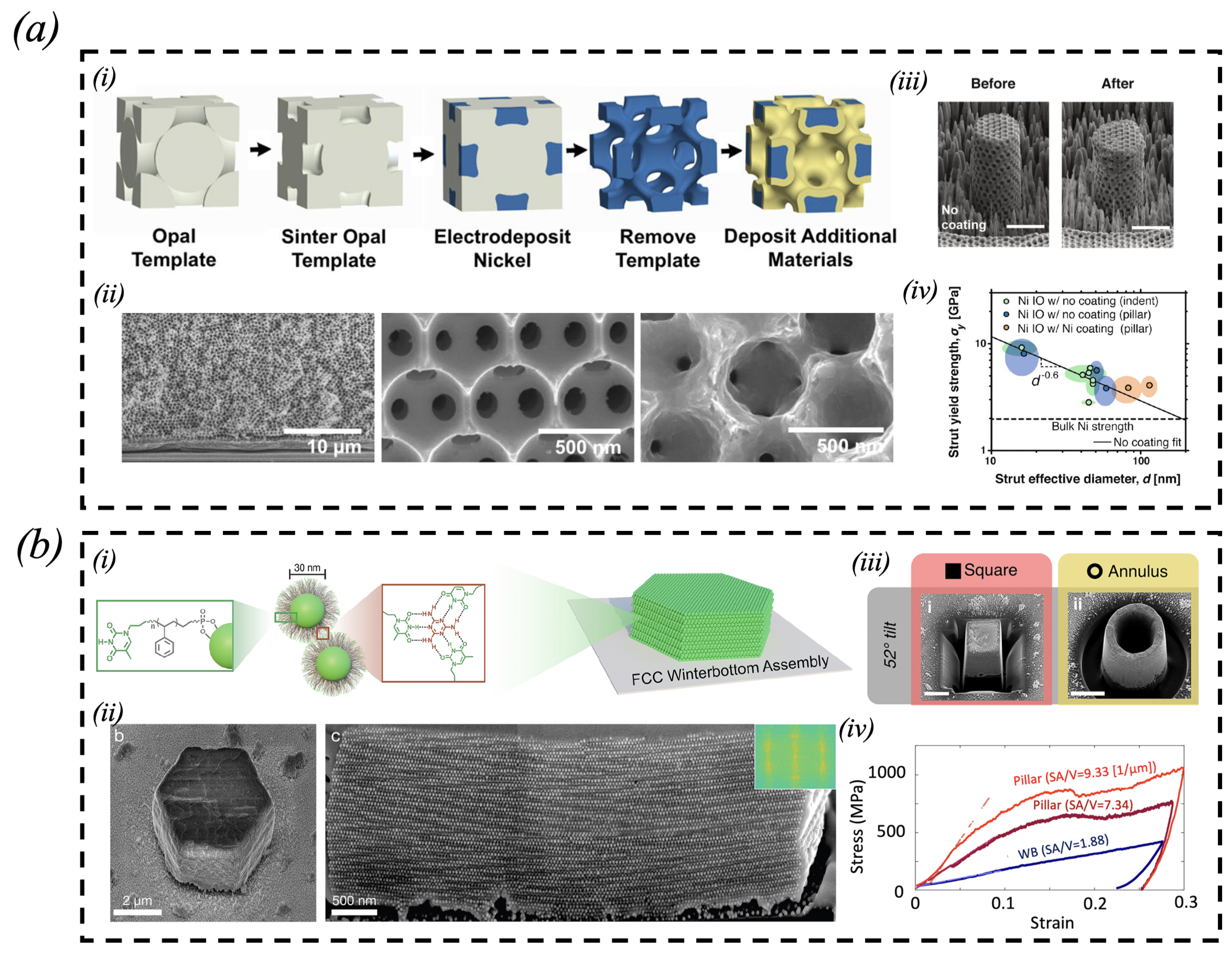}
    \caption{(a) Inverse opals fabricated from colloidal crystal self-assembly: (i) Fabrication process of nickel inverse opals; (ii) Scanning electron microscopy (SEM) images showing cross-sections of nickel inverse opals; (iii) SEM images showing inverse opal micropillars, scale bar is 3 $\mu$m; (iv) Mechanical responses of inverse opal micropillars with different structure diameters (Reprinted with permission from \cite{pikul2019high}. Copyright 2019, the authors). (b) Self-assembled Au superlattices by polystyrene: (ii) Fabrication of Au superlattices; (ii) SEM images of superlattices and cross-sectional view; (iii) Shape morphing of superlattices by FIB, scale bar is 2 $\mu$m; (iv) Tunable mechanical responses of different surface area/volume (SA/V) ratios (Reprinted with permission from \cite{dhulipala2023tunable}. Copyright 2023, ACS).}
    \label{fig:abstract}
\end{figure}

Colloidal nanocrystal self-assembly is another approach to fabricating mechanical metamaterials \cite{boles2016self, zhang2010colloidal,mirkin2023inspired}. \textbf{Fig.1(b)} illustrates the self-assembly processes. Colloidal nanocrystals, with diverse geometries, are first coated with surface ligands such as polymer brushes \cite{ye2015structural} or DNA strands \cite{zhang2013general}. Then, they self-assemble into nano- and micro-structures as a result of interparticle interactions, local geometric constraints, and the surrounding environment \cite{boles2016self}. Like atoms in a crystal lattice, these nanocrystals can autonomously organize into various lattice structures, e.g., face-centered cubic (FCC) or hexagonal close-packed (HCP), depending on the particle geometry and environment. 

Self-assembled colloidal crystals have been used as a template for fabricating mechanical metamaterials. A notable example is inverse opals, which have a 3D periodic nano- or micro-scale porous architecture \cite{hatton2010assembly,marlow2009opals}. As depicted in \textbf{Fig.2(ai)}, the fabrication process involves the self-assembly of colloidal crystals, followed by infiltrating the interstitial spaces with precursor materials like polymers \cite{chen2008fabrication}, ceramics \cite{do2015self}, or metals \cite{pikul2019high}. After removing the colloidal crystals through methods like calcination or dissolution, a three-dimensional continuous phase defined by a uniform network of interconnected voids emerges, as seen in \textbf{Fig.2(aii)}. The resulting materials exhibit superior mechanical properties when compared with bulk phases. For example, porous nickel inverse opals exhibit superior mechanical strength, an attribute controlled by plasticity size effects associated with struct effective diameter (\textbf{Fig.2(aiii)} \& \textbf{(aiv)}), which is determined by the size of the colloidal crystals [49]. However, this multi-step process, which requires the synthesis of colloidal crystals, infiltration, and template removal, can be time-consuming. Potential issues like shrinkage and cracking might occur during the template removal \cite{hatton2010assembly}.

\begin{figure}[!ht]
    \centering
    \includegraphics[width=1.0\textwidth]{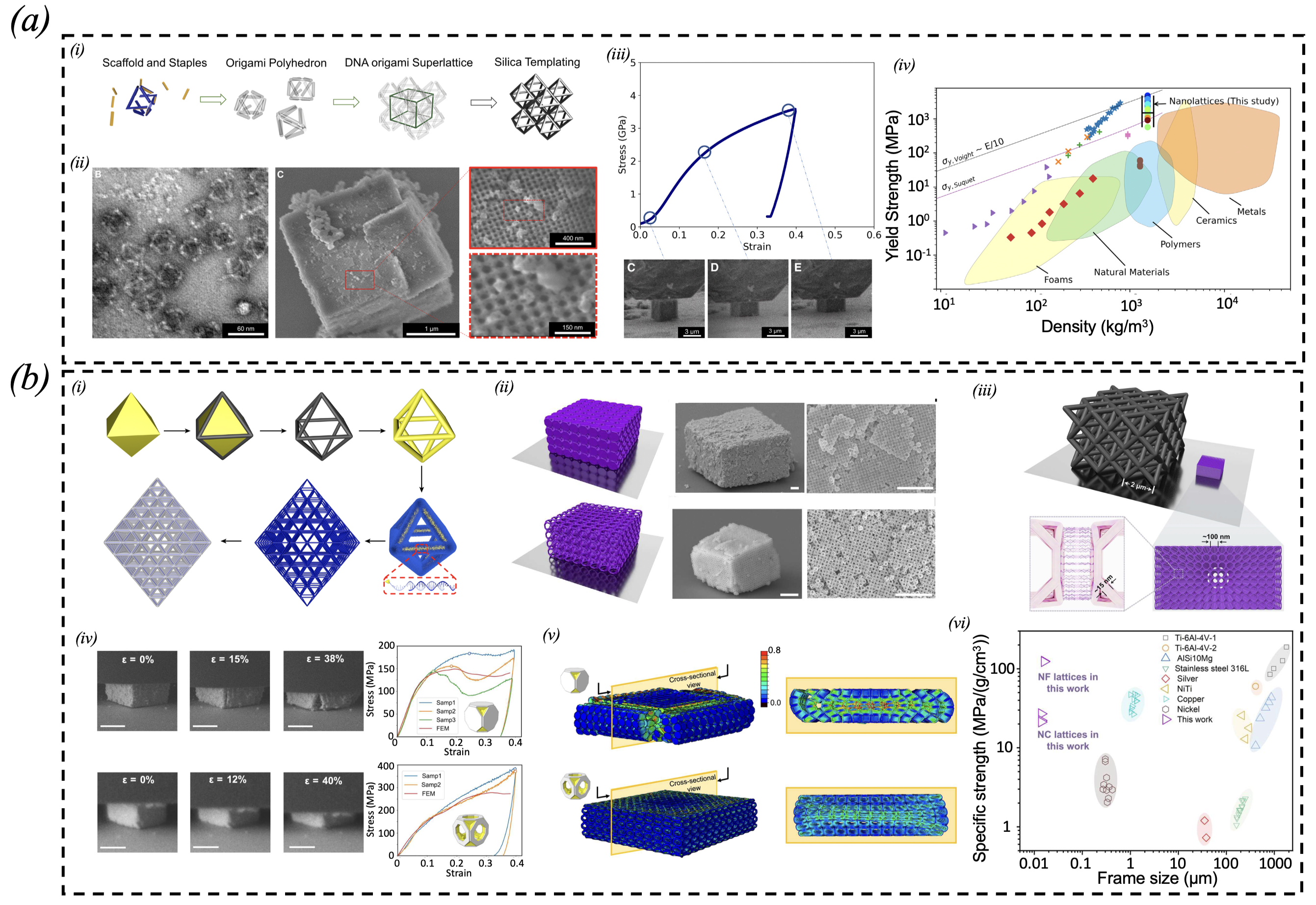}
    \caption{(a) Mechanical metamaterials fabricated from DNA origami techniques: (i) Fabrication process from DNA origami to silica-coated structure; (ii) Transmission electron microscopy (TEM) and SEM images for DNA origami frames and final assembled structures; (iii) In-situ nanomechanical testing; (iv) Ashby diagram with comparison to previously reported lattices (Reprinted with permission from \cite{michelson2023high}. Copyright 2023, the authors). (b) DNA self-assembled ultra-strong colloidal crystal mechanical metamaterials: (i) Fabrication process from open-channel metallic frame to self-assembled metamaterials (Reprinted with permission from \cite{li2022open}. Copyright 2022, Springer Nature); (ii) Schematic and SEM images of investigated metamaterials, scale bar is 1 $\mu$m; (iii) Comparison between TPP printed lattice and self-assembled nanolattices; (iv) In-situ nanomechanical testing, scale bar is 2 $\mu$m; (v) Deformation mechanisms as revealed by FEA; (vi) Feature size vs. specific strength Ashby diagram (Reprinted with permission from \cite{jin2023ultrastrong}. Copyright 2023, the authors). }
    \label{fig:abstract}
\end{figure}

An alternative approach is the direct fabrication of mechanical metamaterial through the self-assembly of strong nanoparticle superlattices \cite{lee2022nanoparticle}. For example, Au nanoparticle superlattices can be fabricated using a DNA self-assembly technique \cite{elghanian1997selective, lewis2020using, mirkin2020dna}. Nanoindentation experiments using an Atomic Force Microscope (AFM) revealed that their stiffness can be modulated by tuning both the DNA length and nanoparticle dimensions \cite{lewis2020using}. More recently, Au nanoparticle superlattices were self-assembled using polystyrene, as shown in  \textbf{Fig.2(bi)} and \textbf{Fig.2(bii)} \cite{dhulipala2023tunable}. When these superlattices were carved into samples with different surface area to volume (SA/V) ratios using focused ion milling, they exhibited tunable mechanical behaviors as revealed by in-situ nanomechanical testing and discrete element method (DEM) simulations (\textbf{Fig.2(biii)} \& \textbf{(biv)}) \cite{dhulipala2023tunable}.

To leverage the distinctive architectural properties of mechanical metamaterials, it is desired to design self-assembled materials with tailored architectures beyond solid nanocrystals. A contemporary strategy employs DNA origami techniques to construct such architectural templates \cite{dey2021dna, castro2011primer,hong2017dna}. For instance, as illustrated in \textbf{Fig.3(ai)}, DNA octahedron frames were synthesized using the DNA origami technique \cite{michelson2023high}. These octahedron frames were then self-assembled into a superlattice via vertex-to-vertex linkages between the frames. Subsequently, silica was templated onto the DNA scaffold, forming a strong nano-architected material with truss diameters ranging from 4 to 20 nm (\textbf{Fig.3(aii)}). The material displayed a superior compressive yield strength between 1 and 5 GPa, as unveiled by in-situ nanomechanical tests, see \textbf{Fig.3(aiii)}. When compressive yield strength and density are represented in an Ashby diagram, \textbf{Fig.3(aiv)}, these DNA-templated silica metamaterials surpass most known engineering materials with similar density. Similar findings are reported in \cite{kulikowski4510528dna}. 

Another synthesis strategy involves the use of open-channel metallic nanocrystals to fabricate metamaterials using DNA self-assembly \cite{li2022open}. As shown in \textbf{Fig.3(bi)}, the metamaterials were fabricated through three key steps: (1) designing truncated cubic Au nanoparticles; (2) undertaking edge-selective or facet-selective growth of Pt on these truncated Au nanoparticles followed by selective etching of Au; and (3) employing DNA to self-assemble these open-channel nanocrystals \cite{li2022open}. These structures, with unit cell sizes of ~100 nm and feature sizes of around 15 nm (\textbf{Fig.3(bii)} \& \textbf{(biii)}), offer high turnability in their topologies and densities, and consequently in their mechanical properties. We performed combined in-situ nanomechanical experiments (\textbf{Fig.3(biv)}) and finite element analysis (FEA) (\textbf{Fig.3(bv)}) to understand their structural-property relations. We found that architected nanolattices possess 6 times higher specific strength compared to solid nanolattices, attributed to the compressibility of unit cells, which can prevent DNA from premature failure, and a smaller is stronger nanoscale material size effects \cite{jin2023ultrastrong}. Remarkably, as shown in \textbf{Fig.3(bvi)}, their specific strength is similar to millimeter-scale lattices produced via AM, emphasizing the potential of self-assembly techniques in pioneering ultra-strong nano-architected mechanical metamaterials with unprecedented resolution.

\section{Perspectives}

In the previous sections, we summarized selected advances in the utilization of self-assembly techniques to fabricate mechanical metamaterials with exceptional mechanical properties, architectural design versatility, and fabrication at scale. These features make the technology particularly suitable for a variety of applications ranging from microelectronics to medical devices to aerospace protective coating. In this section, we delve into current challenges in the field and discuss opportunities for research directions in self-assembled mechanical metamaterials.

\subsection{Main challenges in scaling-up and opportunities in nanomechanics}

While self-assembled mechanical metamaterials exhibit superior mechanical behaviors, scaling up these materials from a laboratory scale to a practical engineering scale remains a formidable challenge. For instance, the 3D copolymer self-assembly technique was demonstrated up to centimeter-scale, yet the nanostructures are restricted by the phase-separation process, limiting the tunability of their mechanical properties. Inspired by hierarchical architectures commonly observed in load-bearing natural materials, we envision opportunities driven by recent developments in macromolecular engineering, particularly sequence-controlled copolymers, which could provide new strategies for coding macromolecules into 3D construct complexes. Furthermore, postprocessing techniques, such as electro-plating metallic coatings on large-scale polymer templates, are likely to produce nonuniform structures. Therefore, refining current nanofabrication methods appears crucial to synthesize spatially uniform and reproducible nanostructures emerging from the copolymer self-assembly process.

In contrast, the colloidal self-assembly technique has been primarily used for micro-scale 3D structures or 2D metasurfaces, such as photonic metamaterials for waveguiding. One of the current challenges in colloidal self-assembly is producing large quantities of high-quality nanocrystals with low polydispersity. As synthesis methods advance, we anticipate that large volumes of nanocrystals will be readily available, including customization of architecture, feature size, and constituents. It is worth mentioning that scaling up isn't solely about producing larger volumes of metamaterials. Synthesis defects and imperfections, such as vacancies, declinations, and/or grain boundaries, could potentially impact their mechanical properties. While producing perfect crystals may be desirable in some applications, it's important to recognize that, like in traditional materials, scaling of dimensions will inevitably result in various defect types that can play a crucial role in material performance. This underscores the need for comprehensive fundamental studies across scales on synthesis-property relationships, including the role of defect engineering on mechanical behaviors.

\subsection{Developing next-generation self-assembly techniques}

Looking forward, we anticipate fabrication technique advances that will result in mechanical metamaterials with unprecedented precision, scalability, and adaptability. One promising direction is the advancement of open-channel colloidal crystal metamaterials, which can serve as a foundational platform because of their versatility \cite{li2022open}. By innovating on unit cell architecture based on mechanics analysis, new structures with superior mechanical performance, insensitivity to defects, and structural integrity can be engineered. Furthermore, dual- or multi-phased open-channel colloidal crystal metamaterials can be designed and synthesized to exhibit mechanical behaviors like sequential buckling. Also, it is desirable to engineer multi-scale colloidal crystal metamaterials with hierarchical structures. Initial self-assembly could first generate a superlattice scaffold, which could then serve as building blocks for a secondary assembly process, creating complex multi-level open-channel architectures. Beyond metallic building blocks, there is potential to integrate other material constituents, like hydrogels or ceramics, depending on the applications. 

Another intriguing prospect is to develop 4D self-assembly techniques, which are advancing the field of “programmable matter”. Similar to contemporary 4D printing techniques \cite{peng20224d,fu20224d, sun2024perspective}, which introduce time as an additional dimension, 4D self-assembly techniques enable functionalized nanocrystals to change their packing geometry, mechanical properties, and hence functionality in response to environmental stimuli such as far-field pressure \cite{jin2021ruga}, light \cite{leanza2023active}, or magnetic field \cite{kim2018printing} are envisioned. Such dynamic response could be achieved through the integration of phase-changing constituents, e.g., stimuli-responsive polymers or embedded microactuators. Such capability will enable self-assembled metamaterial to transition between various programable mechanical states. 

To synthetize multi-phase metamaterials, hybrid fabrication approaches combining top-down methods (like photolithography or AM) with bottom-up self-assembly can be employed \cite{sola2023self}. For instance, macro-scale templates fabricated from AM could be used to guide the self-assembly at micro- or nanoscales. For example, the combination of 3D bioprinting and DNA nanotechnology could fabricate bio-architected materials with resolutions at the sub-cellular level, thus paving the way for innovative bio-synthetic systems and applications \cite{jakel2023multiscale}. Moreover, it is desirable that next-generation metamaterials be produced sustainably. For instance, by leveraging bio-derived building blocks and green chemistry, advanced hybrid fabrication processes can provide sustainable routes to mass-produce green metamaterials.

\subsection{Developing new characterization methods}

Characterizing mechanical metamaterials requires advanced techniques that can provide insights into their structure-property relations across multiple scales. As these metamaterials grow in complexity, developing new experimental and computational techniques is crucial. Here are some new techniques we think could address future characterization needs. First, to understand the complex self-assembly processes and their mechanical properties, techniques that allow real-time observation under nanoindentation or nano-compression are needed. Integrating high-resolution microscopy, such as TEM and SEM, with miniaturized mechanical testing stages would be invaluable to understanding their deformation mechanisms \cite{jin2023ultrastrong, ramachandramoorthy2015pushing, bhowmick2019advanced}. Additionally, techniques like in-situ X-ray micro- or nano-tomography can offer comprehensive structural insights into these metamaterials, helping to correlate nanostructures with mechanical properties \cite{shaikeea2022toughness}. Combining in-situ nanomechanical testing with Raman and 3D tomography can provide concurrent insights into their mechanical, chemical, and structural aspects. Another opportunity in mechanical characterization is the development of high-throughput testing. Given the small size of these self-assembled crystals, manual identification of samples and data acquisition can be labor intensive. Hence, developing automated and high throughput testing platforms would expedite sample identification and mechanical testing. Techniques we created to automate single cell manipulation and analysis appear promising \cite{mukherjee2022deep, patino2022multiplexed}. Moreover, such experiments would also provide a rich dataset for an inverse design workflow of metamaterials.

Besides experimental characterization, multiscale simulations are needed to uncover their self-assembly during synthesis, as well as the emergent properties of synthesized materials. For example, molecular dynamics (MD) can provide insights into self-assembly processes under different environmental conditions \cite{samanta2019multivalent}. Furthermore, MD can also help extract constituent properties, e.g., the cohesive law of DNA-assembled interfaces, facilitating subsequent FEA simulations designed to reveal deformation and failure modes at the structural level. As in other applications of multiscale modeling, challenges arising from differences in time scales between MD and experiments need to be accounted for \cite{hollingsworth2018molecular}. While multiscale computational simulations offer valuable insights into metamaterial designs, they can be time-consuming. Hence, other rapidly evolving numerical methods, such as the direct element method (DEM) or coarse-grained MD, should be considered.

\subsection{Machine learning assisted inverse design}

Currently, the design of self-assembled mechanical metamaterials is primarily a forward- process based on laboratory trial-and-error. In our view, scientific machine learning (SciML), together with advanced experimental characterization and computational modeling, will play a pivotal role in designing and optimizing future self-assembled mechanical metamaterials \cite{jin2023recent}. ML has already shown potential in the inverse design of mechanical metamaterials from AM \cite{guo2021artificial}. This task becomes inherently complex and computationally challenging for metamaterials synthesized by self-assembly, where the interplay of interparticle interactions and the effect of defects become critical, rendering the design space hyper-dimensional and non-intuitive. The hyper-dimensionality challenge calls for the utilization of unsupervised ML algorithms, e.g., principal component analysis and clustering, to narrow down the design space to the most important variables \cite{alderete2022machine}. Likewise, the use of symbolic regression for dimensionality reduction and extraction of governing laws from experiments appears promising \cite{schmidt2009distilling}. Then, a more manageable number of simulations and experiments will be needed to optimally design and validate solutions. In this respect, a variety of neural network architectures can be combined with other optimization methods like Genetic Algorithm (GA) to select the most desirable designs \cite{mirjalili2019genetic}. In years to come, we envision the utilization of SciML like physics-informed neural networks\cite{karniadakis2021physics, raissi2019physics, song2023identifying, zhang2022analyses} or neural operators \cite{lu2021learning,goswami2023physics}, integrated with automated labs, to design and optimize self-assembled metamaterials in real time for desired multifunctional properties. The inverse design of mechanical metamaterials with desired mechanical performance will not only assist in the selection of material constituents, unit cell geometry, and self-assembly environments but also provide insights into governing principles through ML interpretability \cite{krenn2022scientific}, an emergent research topic.

\section{Conclusions}
Self-assembly offers a potentially scalable route to fabricate mechanical metamaterials with unprecedented properties. In this perspective, we first provide an overview of self-assembly techniques for mechanical metamaterials, with a focus on copolymer and colloid crystal self-assembly processes. Then, we provide our perspectives in this field, spanning fabrication, characterization, and ML-assisted inverse design. We hope this perspective could provide valuable insights for researchers in this field to design, fabricate, and characterize next-generation mechanical metamaterials from self-assembly. As this field undergoes rapid evolution, fostering collaborations and building bridges between disciplines such as materials science, computation mechanics, and chemistry becomes necessary to address challenges and opportunities. Likewise, inspiration from other scientific endeavors, e.g., cell biology and medicine, which are achieving great success based on the merger of lab automation and ML, seems very promising. 

\section*{Acknowledgment}
H.D.E. acknowledges the financial support from the Air Force Office of Scientific Research (AFOSR-FA9550-20-1-0258), National Science Foundation (grant CMMI-1953806), and Office of Naval Research (grant N000142212133).

\section*{Author Contribution}
H.J. and H.D.E. conceived the initial idea to write this perspective and defined its content. H.J. wrote the initial draft. H.D.E edited the manuscript. All authors gave final approval for publication.

\bibliographystyle{elsarticle-num-names}
\bibliography{reference.bib}

\begin{thebibliography}{88}
\expandafter\ifx\csname natexlab\endcsname\relax\def\natexlab#1{#1}\fi
\providecommand{\url}[1]{\texttt{#1}}
\providecommand{\href}[2]{#2}
\providecommand{\path}[1]{#1}
\providecommand{\DOIprefix}{doi:}
\providecommand{\ArXivprefix}{arXiv:}
\providecommand{\URLprefix}{URL: }
\providecommand{\Pubmedprefix}{pmid:}
\providecommand{\doi}[1]{\href{http://dx.doi.org/#1}{\path{#1}}}
\providecommand{\Pubmed}[1]{\href{pmid:#1}{\path{#1}}}
\providecommand{\bibinfo}[2]{#2}
\ifx\xfnm\relax \def\xfnm[#1]{\unskip,\space#1}\fi
\bibitem[{Xia et~al.(2022)Xia, Spadaccini, and Greer}]{xia2022responsive}
\bibinfo{author}{X.~Xia}, \bibinfo{author}{C.~M. Spadaccini}, \bibinfo{author}{J.~R. Greer},
\newblock \bibinfo{title}{Responsive materials architected in space and time},
\newblock \bibinfo{journal}{Nature Reviews Materials} \bibinfo{volume}{7} (\bibinfo{year}{2022}) \bibinfo{pages}{683--701}.
\bibitem[{Montemayor et~al.(2015)Montemayor, Chernow, and Greer}]{montemayor2015materials}
\bibinfo{author}{L.~Montemayor}, \bibinfo{author}{V.~Chernow}, \bibinfo{author}{J.~R. Greer},
\newblock \bibinfo{title}{Materials by design: Using architecture in material design to reach new property spaces},
\newblock \bibinfo{journal}{Mrs Bulletin} \bibinfo{volume}{40} (\bibinfo{year}{2015}) \bibinfo{pages}{1122--1129}.
\bibitem[{Bauer et~al.(2017)Bauer, Meza, Schaedler, Schwaiger, Zheng, and Valdevit}]{bauer2017nanolattices}
\bibinfo{author}{J.~Bauer}, \bibinfo{author}{L.~R. Meza}, \bibinfo{author}{T.~A. Schaedler}, \bibinfo{author}{R.~Schwaiger}, \bibinfo{author}{X.~Zheng}, \bibinfo{author}{L.~Valdevit},
\newblock \bibinfo{title}{Nanolattices: an emerging class of mechanical metamaterials},
\newblock \bibinfo{journal}{Advanced Materials} \bibinfo{volume}{29} (\bibinfo{year}{2017}) \bibinfo{pages}{1701850}.
\bibitem[{Surjadi et~al.(2019)Surjadi, Gao, Du, Li, Xiong, Fang, and Lu}]{surjadi2019mechanical}
\bibinfo{author}{J.~U. Surjadi}, \bibinfo{author}{L.~Gao}, \bibinfo{author}{H.~Du}, \bibinfo{author}{X.~Li}, \bibinfo{author}{X.~Xiong}, \bibinfo{author}{N.~X. Fang}, \bibinfo{author}{Y.~Lu},
\newblock \bibinfo{title}{Mechanical metamaterials and their engineering applications},
\newblock \bibinfo{journal}{Advanced Engineering Materials} \bibinfo{volume}{21} (\bibinfo{year}{2019}) \bibinfo{pages}{1800864}.
\bibitem[{Zhang et~al.(2019)Zhang, Vyatskikh, Gao, Greer, and Li}]{zhang2019lightweight}
\bibinfo{author}{X.~Zhang}, \bibinfo{author}{A.~Vyatskikh}, \bibinfo{author}{H.~Gao}, \bibinfo{author}{J.~R. Greer}, \bibinfo{author}{X.~Li},
\newblock \bibinfo{title}{Lightweight, flaw-tolerant, and ultrastrong nanoarchitected carbon},
\newblock \bibinfo{journal}{Proceedings of the National Academy of Sciences} \bibinfo{volume}{116} (\bibinfo{year}{2019}) \bibinfo{pages}{6665--6672}.
\bibitem[{Li* et~al.(2023)Li*, Jin*, Zhou, Wang, Lin, Mirkin, and Espinosa}]{jin2023ultrastrong}
\bibinfo{author}{Y.~Li*}, \bibinfo{author}{H.~Jin*}, \bibinfo{author}{W.~Zhou}, \bibinfo{author}{Z.~Wang}, \bibinfo{author}{Z.~Lin}, \bibinfo{author}{C.~A. Mirkin}, \bibinfo{author}{H.~D. Espinosa},
\newblock \bibinfo{title}{Ultrastrong colloidal crystal metamaterials engineered with dna},
\newblock \bibinfo{journal}{Science Advances} \bibinfo{volume}{9} (\bibinfo{year}{2023}) \bibinfo{pages}{eadj8103}.
\bibitem[{Meza et~al.(2014)Meza, Das, and Greer}]{meza2014strong}
\bibinfo{author}{L.~R. Meza}, \bibinfo{author}{S.~Das}, \bibinfo{author}{J.~R. Greer},
\newblock \bibinfo{title}{Strong, lightweight, and recoverable three-dimensional ceramic nanolattices},
\newblock \bibinfo{journal}{Science} \bibinfo{volume}{345} (\bibinfo{year}{2014}) \bibinfo{pages}{1322--1326}.
\bibitem[{Lin et~al.(2020)Lin, Novelino, Wei, Alderete, Paulino, Espinosa, and Krishnaswamy}]{lin2020folding}
\bibinfo{author}{Z.~Lin}, \bibinfo{author}{L.~S. Novelino}, \bibinfo{author}{H.~Wei}, \bibinfo{author}{N.~A. Alderete}, \bibinfo{author}{G.~H. Paulino}, \bibinfo{author}{H.~D. Espinosa}, \bibinfo{author}{S.~Krishnaswamy},
\newblock \bibinfo{title}{Folding at the microscale: Enabling multifunctional 3d origami-architected metamaterials},
\newblock \bibinfo{journal}{Small} \bibinfo{volume}{16} (\bibinfo{year}{2020}) \bibinfo{pages}{2002229}.
\bibitem[{Portela et~al.(2021)Portela, Edwards, Veysset, Sun, Nelson, Kochmann, and Greer}]{portela2021supersonic}
\bibinfo{author}{C.~M. Portela}, \bibinfo{author}{B.~W. Edwards}, \bibinfo{author}{D.~Veysset}, \bibinfo{author}{Y.~Sun}, \bibinfo{author}{K.~A. Nelson}, \bibinfo{author}{D.~M. Kochmann}, \bibinfo{author}{J.~R. Greer},
\newblock \bibinfo{title}{Supersonic impact resilience of nanoarchitected carbon},
\newblock \bibinfo{journal}{Nature Materials} \bibinfo{volume}{20} (\bibinfo{year}{2021}) \bibinfo{pages}{1491--1497}.
\bibitem[{Vyatskikh et~al.(2018)Vyatskikh, Delalande, Kudo, Zhang, Portela, and Greer}]{vyatskikh2018additive}
\bibinfo{author}{A.~Vyatskikh}, \bibinfo{author}{S.~Delalande}, \bibinfo{author}{A.~Kudo}, \bibinfo{author}{X.~Zhang}, \bibinfo{author}{C.~M. Portela}, \bibinfo{author}{J.~R. Greer},
\newblock \bibinfo{title}{Additive manufacturing of 3d nano-architected metals},
\newblock \bibinfo{journal}{Nature communications} \bibinfo{volume}{9} (\bibinfo{year}{2018}) \bibinfo{pages}{593}.
\bibitem[{Gibson et~al.(2021)Gibson, Rosen, Stucker, Khorasani, Rosen, Stucker, and Khorasani}]{gibson2021additive}
\bibinfo{author}{I.~Gibson}, \bibinfo{author}{D.~W. Rosen}, \bibinfo{author}{B.~Stucker}, \bibinfo{author}{M.~Khorasani}, \bibinfo{author}{D.~Rosen}, \bibinfo{author}{B.~Stucker}, \bibinfo{author}{M.~Khorasani}, \bibinfo{title}{Additive manufacturing technologies}, volume~\bibinfo{volume}{17}, \bibinfo{publisher}{Springer}, \bibinfo{year}{2021}.
\bibitem[{Frazier(2014)}]{frazier2014metal}
\bibinfo{author}{W.~E. Frazier},
\newblock \bibinfo{title}{Metal additive manufacturing: a review},
\newblock \bibinfo{journal}{Journal of Materials Engineering and performance} \bibinfo{volume}{23} (\bibinfo{year}{2014}) \bibinfo{pages}{1917--1928}.
\bibitem[{Chen et~al.(2022)Chen, Van~Petegem, Zou, Simonelli, Tse, Chang, Makowska, Sanchez, and Moens-Van~Swygenhoven}]{chen2022microstructural}
\bibinfo{author}{M.~Chen}, \bibinfo{author}{S.~Van~Petegem}, \bibinfo{author}{Z.~Zou}, \bibinfo{author}{M.~Simonelli}, \bibinfo{author}{Y.~Y. Tse}, \bibinfo{author}{C.~S.~T. Chang}, \bibinfo{author}{M.~G. Makowska}, \bibinfo{author}{D.~F. Sanchez}, \bibinfo{author}{H.~Moens-Van~Swygenhoven},
\newblock \bibinfo{title}{Microstructural engineering of a dual-phase ti-al-v-fe alloy via in situ alloying during laser powder bed fusion},
\newblock \bibinfo{journal}{Additive Manufacturing} \bibinfo{volume}{59} (\bibinfo{year}{2022}) \bibinfo{pages}{103173}.
\bibitem[{Wong and Hernandez(2012)}]{wong2012review}
\bibinfo{author}{K.~V. Wong}, \bibinfo{author}{A.~Hernandez},
\newblock \bibinfo{title}{A review of additive manufacturing},
\newblock \bibinfo{journal}{International scholarly research notices} \bibinfo{volume}{2012} (\bibinfo{year}{2012}).
\bibitem[{O'Halloran et~al.(2023)O'Halloran, Pandit, Heise, and Kellett}]{o2023two}
\bibinfo{author}{S.~O'Halloran}, \bibinfo{author}{A.~Pandit}, \bibinfo{author}{A.~Heise}, \bibinfo{author}{A.~Kellett},
\newblock \bibinfo{title}{Two-photon polymerization: Fundamentals, materials, and chemical modification strategies},
\newblock \bibinfo{journal}{Advanced Science} \bibinfo{volume}{10} (\bibinfo{year}{2023}) \bibinfo{pages}{2204072}.
\bibitem[{Carlotti and Mattoli(2019)}]{carlotti2019functional}
\bibinfo{author}{M.~Carlotti}, \bibinfo{author}{V.~Mattoli},
\newblock \bibinfo{title}{Functional materials for two-photon polymerization in microfabrication},
\newblock \bibinfo{journal}{Small} \bibinfo{volume}{15} (\bibinfo{year}{2019}) \bibinfo{pages}{1902687}.
\bibitem[{Mai and Eisenberg(2012)}]{mai2012self}
\bibinfo{author}{Y.~Mai}, \bibinfo{author}{A.~Eisenberg},
\newblock \bibinfo{title}{Self-assembly of block copolymers},
\newblock \bibinfo{journal}{Chemical Society Reviews} \bibinfo{volume}{41} (\bibinfo{year}{2012}) \bibinfo{pages}{5969--5985}.
\bibitem[{Grzelczak et~al.(2010)Grzelczak, Vermant, Furst, and Liz-Marz{\'a}n}]{grzelczak2010directed}
\bibinfo{author}{M.~Grzelczak}, \bibinfo{author}{J.~Vermant}, \bibinfo{author}{E.~M. Furst}, \bibinfo{author}{L.~M. Liz-Marz{\'a}n},
\newblock \bibinfo{title}{Directed self-assembly of nanoparticles},
\newblock \bibinfo{journal}{ACS nano} \bibinfo{volume}{4} (\bibinfo{year}{2010}) \bibinfo{pages}{3591--3605}.
\bibitem[{Cai et~al.(2021)Cai, Li, Ravaine, He, Song, Yin, Zheng, Teng, and Zhang}]{cai2021colloidal}
\bibinfo{author}{Z.~Cai}, \bibinfo{author}{Z.~Li}, \bibinfo{author}{S.~Ravaine}, \bibinfo{author}{M.~He}, \bibinfo{author}{Y.~Song}, \bibinfo{author}{Y.~Yin}, \bibinfo{author}{H.~Zheng}, \bibinfo{author}{J.~Teng}, \bibinfo{author}{A.~Zhang},
\newblock \bibinfo{title}{From colloidal particles to photonic crystals: Advances in self-assembly and their emerging applications},
\newblock \bibinfo{journal}{Chemical Society Reviews} \bibinfo{volume}{50} (\bibinfo{year}{2021}) \bibinfo{pages}{5898--5951}.
\bibitem[{Vignolini et~al.(2012)Vignolini, Yufa, Cunha, Guldin, Rushkin, Stefik, Hur, Wiesner, Baumberg, and Steiner}]{vignolini20123d}
\bibinfo{author}{S.~Vignolini}, \bibinfo{author}{N.~A. Yufa}, \bibinfo{author}{P.~S. Cunha}, \bibinfo{author}{S.~Guldin}, \bibinfo{author}{I.~Rushkin}, \bibinfo{author}{M.~Stefik}, \bibinfo{author}{K.~Hur}, \bibinfo{author}{U.~Wiesner}, \bibinfo{author}{J.~J. Baumberg}, \bibinfo{author}{U.~Steiner},
\newblock \bibinfo{title}{A 3d optical metamaterial made by self-assembly},
\newblock \bibinfo{journal}{Advanced Materials} \bibinfo{volume}{24} (\bibinfo{year}{2012}) \bibinfo{pages}{OP23--OP27}.
\bibitem[{Wang et~al.(2016)Wang, Gaitanaros, Lee, Bathe, Shih, and Ke}]{wang2016programming}
\bibinfo{author}{P.~Wang}, \bibinfo{author}{S.~Gaitanaros}, \bibinfo{author}{S.~Lee}, \bibinfo{author}{M.~Bathe}, \bibinfo{author}{W.~M. Shih}, \bibinfo{author}{Y.~Ke},
\newblock \bibinfo{title}{Programming self-assembly of dna origami honeycomb two-dimensional lattices and plasmonic metamaterials},
\newblock \bibinfo{journal}{Journal of the American Chemical Society} \bibinfo{volume}{138} (\bibinfo{year}{2016}) \bibinfo{pages}{7733--7740}.
\bibitem[{Whitesides and Grzybowski(2002)}]{whitesides2002self}
\bibinfo{author}{G.~M. Whitesides}, \bibinfo{author}{B.~Grzybowski},
\newblock \bibinfo{title}{Self-assembly at all scales},
\newblock \bibinfo{journal}{Science} \bibinfo{volume}{295} (\bibinfo{year}{2002}) \bibinfo{pages}{2418--2421}.
\bibitem[{Bates and Fredrickson(1999)}]{bates1999block}
\bibinfo{author}{F.~S. Bates}, \bibinfo{author}{G.~H. Fredrickson},
\newblock \bibinfo{title}{Block copolymers—designer soft materials},
\newblock \bibinfo{journal}{Physics today} \bibinfo{volume}{52} (\bibinfo{year}{1999}) \bibinfo{pages}{32--38}.
\bibitem[{Cummins et~al.(2020)Cummins, Lundy, Walsh, Ponsinet, Fleury, and Morris}]{cummins2020enabling}
\bibinfo{author}{C.~Cummins}, \bibinfo{author}{R.~Lundy}, \bibinfo{author}{J.~J. Walsh}, \bibinfo{author}{V.~Ponsinet}, \bibinfo{author}{G.~Fleury}, \bibinfo{author}{M.~A. Morris},
\newblock \bibinfo{title}{Enabling future nanomanufacturing through block copolymer self-assembly: A review},
\newblock \bibinfo{journal}{Nano Today} \bibinfo{volume}{35} (\bibinfo{year}{2020}) \bibinfo{pages}{100936}.
\bibitem[{Boles et~al.(2016)Boles, Engel, and Talapin}]{boles2016self}
\bibinfo{author}{M.~A. Boles}, \bibinfo{author}{M.~Engel}, \bibinfo{author}{D.~V. Talapin},
\newblock \bibinfo{title}{Self-assembly of colloidal nanocrystals: From intricate structures to functional materials},
\newblock \bibinfo{journal}{Chemical reviews} \bibinfo{volume}{116} (\bibinfo{year}{2016}) \bibinfo{pages}{11220--11289}.
\bibitem[{Young and Lovell(2011)}]{young2011introduction}
\bibinfo{author}{R.~J. Young}, \bibinfo{author}{P.~A. Lovell}, \bibinfo{title}{Introduction to polymers}, \bibinfo{publisher}{CRC press}, \bibinfo{year}{2011}.
\bibitem[{Bates and Fredrickson(1990)}]{bates1990block}
\bibinfo{author}{F.~S. Bates}, \bibinfo{author}{G.~H. Fredrickson},
\newblock \bibinfo{title}{Block copolymer thermodynamics: theory and experiment},
\newblock \bibinfo{journal}{Annual review of physical chemistry} \bibinfo{volume}{41} (\bibinfo{year}{1990}) \bibinfo{pages}{525--557}.
\bibitem[{Jin et~al.(2022{\natexlab{a}})Jin, Jiao, Clifton, and Kim}]{jin2022dynamic}
\bibinfo{author}{H.~Jin}, \bibinfo{author}{T.~Jiao}, \bibinfo{author}{R.~J. Clifton}, \bibinfo{author}{K.-S. Kim},
\newblock \bibinfo{title}{Dynamic fracture of a bicontinuously nanostructured copolymer: A deep-learning analysis of big-data-generating experiment},
\newblock \bibinfo{journal}{Journal of the Mechanics and Physics of Solids} \bibinfo{volume}{164} (\bibinfo{year}{2022}{\natexlab{a}}) \bibinfo{pages}{104898}.
\bibitem[{Jin et~al.(2022{\natexlab{b}})Jin, Machnicki, Hegarty, Clifton, and Kim}]{jin2022understanding}
\bibinfo{author}{H.~Jin}, \bibinfo{author}{C.~Machnicki}, \bibinfo{author}{J.~Hegarty}, \bibinfo{author}{R.~J. Clifton}, \bibinfo{author}{K.-S. Kim},
\newblock \bibinfo{title}{Understanding the nanoscale deformation mechanisms of polyurea from in situ afm tensile experiments},
\newblock in: \bibinfo{booktitle}{Challenges in Mechanics of Time Dependent Materials, Mechanics of Biological Systems and Materials \& Micro-and Nanomechanics, Volume 2: Proceedings of the 2021 Annual Conference \& Exposition on Experimental and Applied Mechanics}, \bibinfo{organization}{Springer}, \bibinfo{year}{2022}{\natexlab{b}}, pp. \bibinfo{pages}{45--51}.
\bibitem[{Jin(2022)}]{jin2022big}
\bibinfo{author}{H.~Jin}, \bibinfo{title}{Big-data-driven multi-scale experimental study of nanostructured block copolymer’s dynamic toughness}, Ph.D. thesis, Brown University, \bibinfo{year}{2022}.
\bibitem[{Kim et~al.(2021)Kim, Jin, Jiao, and Clifton}]{kim2021dynamic}
\bibinfo{author}{K.-S. Kim}, \bibinfo{author}{H.~Jin}, \bibinfo{author}{T.~Jiao}, \bibinfo{author}{R.~J. Clifton},
\newblock \bibinfo{title}{Dynamic fracture-toughness testing of a hierarchically nano-structured solid},
\newblock in: \bibinfo{booktitle}{Fracture, Fatigue, Failure and Damage Evolution, Volume 3: Proceedings of the 2020 Annual Conference on Experimental and Applied Mechanics}, \bibinfo{organization}{Springer}, \bibinfo{year}{2021}, pp. \bibinfo{pages}{97--100}.
\bibitem[{Flory(1942)}]{flory1942thermodynamics}
\bibinfo{author}{P.~J. Flory},
\newblock \bibinfo{title}{Thermodynamics of high polymer solutions},
\newblock \bibinfo{journal}{The Journal of chemical physics} \bibinfo{volume}{10} (\bibinfo{year}{1942}) \bibinfo{pages}{51--61}.
\bibitem[{Huggins(1941)}]{huggins1941solutions}
\bibinfo{author}{M.~L. Huggins},
\newblock \bibinfo{title}{Solutions of long chain compounds},
\newblock \bibinfo{journal}{The Journal of chemical physics} \bibinfo{volume}{9} (\bibinfo{year}{1941}) \bibinfo{pages}{440--440}.
\bibitem[{Hu et~al.(2014)Hu, Gopinadhan, and Osuji}]{hu2014directed}
\bibinfo{author}{H.~Hu}, \bibinfo{author}{M.~Gopinadhan}, \bibinfo{author}{C.~O. Osuji},
\newblock \bibinfo{title}{Directed self-assembly of block copolymers: a tutorial review of strategies for enabling nanotechnology with soft matter},
\newblock \bibinfo{journal}{Soft matter} \bibinfo{volume}{10} (\bibinfo{year}{2014}) \bibinfo{pages}{3867--3889}.
\bibitem[{Park et~al.(2003)Park, Yoon, and Thomas}]{park2003enabling}
\bibinfo{author}{C.~Park}, \bibinfo{author}{J.~Yoon}, \bibinfo{author}{E.~L. Thomas},
\newblock \bibinfo{title}{Enabling nanotechnology with self assembled block copolymer patterns},
\newblock \bibinfo{journal}{Polymer} \bibinfo{volume}{44} (\bibinfo{year}{2003}) \bibinfo{pages}{6725--6760}.
\bibitem[{Vukovic et~al.(2013)Vukovic, ten Brinke, and Loos}]{vukovic2013block}
\bibinfo{author}{I.~Vukovic}, \bibinfo{author}{G.~ten Brinke}, \bibinfo{author}{K.~Loos},
\newblock \bibinfo{title}{Block copolymer template-directed synthesis of well-ordered metallic nanostructures},
\newblock \bibinfo{journal}{Polymer} \bibinfo{volume}{54} (\bibinfo{year}{2013}) \bibinfo{pages}{2591--2605}.
\bibitem[{Khaderi et~al.(2017)Khaderi, Scherer, Hall, Steiner, Ramamurty, Fleck, and Deshpande}]{khaderi2017indentation}
\bibinfo{author}{S.~N. Khaderi}, \bibinfo{author}{M.~Scherer}, \bibinfo{author}{C.~Hall}, \bibinfo{author}{U.~Steiner}, \bibinfo{author}{U.~Ramamurty}, \bibinfo{author}{N.~Fleck}, \bibinfo{author}{V.~Deshpande},
\newblock \bibinfo{title}{The indentation response of nickel nano double gyroid lattices},
\newblock \bibinfo{journal}{Extreme Mechanics Letters} \bibinfo{volume}{10} (\bibinfo{year}{2017}) \bibinfo{pages}{15--23}.
\bibitem[{Filleter et~al.(2012)Filleter, Ryu, Kang, Yin, Bernal, Sohn, Li, Huang, Cai, and Espinosa}]{filleter2012nucleation}
\bibinfo{author}{T.~Filleter}, \bibinfo{author}{S.~Ryu}, \bibinfo{author}{K.~Kang}, \bibinfo{author}{J.~Yin}, \bibinfo{author}{R.~A. Bernal}, \bibinfo{author}{K.~Sohn}, \bibinfo{author}{S.~Li}, \bibinfo{author}{J.~Huang}, \bibinfo{author}{W.~Cai}, \bibinfo{author}{H.~D. Espinosa},
\newblock \bibinfo{title}{Nucleation-controlled distributed plasticity in penta-twinned silver nanowires},
\newblock \bibinfo{journal}{Small} \bibinfo{volume}{8} (\bibinfo{year}{2012}) \bibinfo{pages}{2986--2993}.
\bibitem[{Bernal et~al.(2015)Bernal, Aghaei, Lee, Ryu, Sohn, Huang, Cai, and Espinosa}]{bernal2015intrinsic}
\bibinfo{author}{R.~A. Bernal}, \bibinfo{author}{A.~Aghaei}, \bibinfo{author}{S.~Lee}, \bibinfo{author}{S.~Ryu}, \bibinfo{author}{K.~Sohn}, \bibinfo{author}{J.~Huang}, \bibinfo{author}{W.~Cai}, \bibinfo{author}{H.~Espinosa},
\newblock \bibinfo{title}{Intrinsic bauschinger effect and recoverable plasticity in pentatwinned silver nanowires tested in tension},
\newblock \bibinfo{journal}{Nano letters} \bibinfo{volume}{15} (\bibinfo{year}{2015}) \bibinfo{pages}{139--146}.
\bibitem[{Silverstein(2014)}]{silverstein2014emulsion}
\bibinfo{author}{M.~S. Silverstein},
\newblock \bibinfo{title}{Emulsion-templated porous polymers: A retrospective perspective},
\newblock \bibinfo{journal}{Polymer} \bibinfo{volume}{55} (\bibinfo{year}{2014}) \bibinfo{pages}{304--320}.
\bibitem[{Portela et~al.(2020)Portela, Vidyasagar, Kr{\"o}del, Weissenbach, Yee, Greer, and Kochmann}]{portela2020extreme}
\bibinfo{author}{C.~M. Portela}, \bibinfo{author}{A.~Vidyasagar}, \bibinfo{author}{S.~Kr{\"o}del}, \bibinfo{author}{T.~Weissenbach}, \bibinfo{author}{D.~W. Yee}, \bibinfo{author}{J.~R. Greer}, \bibinfo{author}{D.~M. Kochmann},
\newblock \bibinfo{title}{Extreme mechanical resilience of self-assembled nanolabyrinthine materials},
\newblock \bibinfo{journal}{Proceedings of the National Academy of Sciences} \bibinfo{volume}{117} (\bibinfo{year}{2020}) \bibinfo{pages}{5686--5693}.
\bibitem[{Pikul et~al.(2019)Pikul, {\"O}zerin{\c{c}}, Liu, Zhang, Braun, Deshpande, and King}]{pikul2019high}
\bibinfo{author}{J.~H. Pikul}, \bibinfo{author}{S.~{\"O}zerin{\c{c}}}, \bibinfo{author}{B.~Liu}, \bibinfo{author}{R.~Zhang}, \bibinfo{author}{P.~V. Braun}, \bibinfo{author}{V.~S. Deshpande}, \bibinfo{author}{W.~P. King},
\newblock \bibinfo{title}{High strength metallic wood from nanostructured nickel inverse opal materials},
\newblock \bibinfo{journal}{Scientific reports} \bibinfo{volume}{9} (\bibinfo{year}{2019}) \bibinfo{pages}{719}.
\bibitem[{Dhulipala et~al.(2023)Dhulipala, Yee, Zhou, Sun, Andrade, Macfarlane, and Portela}]{dhulipala2023tunable}
\bibinfo{author}{S.~Dhulipala}, \bibinfo{author}{D.~W. Yee}, \bibinfo{author}{Z.~Zhou}, \bibinfo{author}{R.~Sun}, \bibinfo{author}{J.~E. Andrade}, \bibinfo{author}{R.~J. Macfarlane}, \bibinfo{author}{C.~M. Portela},
\newblock \bibinfo{title}{Tunable mechanical response of self-assembled nanoparticle superlattices},
\newblock \bibinfo{journal}{Nano Letters}  (\bibinfo{year}{2023}).
\bibitem[{Zhang et~al.(2010)Zhang, Li, Zhang, and Yang}]{zhang2010colloidal}
\bibinfo{author}{J.~Zhang}, \bibinfo{author}{Y.~Li}, \bibinfo{author}{X.~Zhang}, \bibinfo{author}{B.~Yang},
\newblock \bibinfo{title}{Colloidal self-assembly meets nanofabrication: From two-dimensional colloidal crystals to nanostructure arrays},
\newblock \bibinfo{journal}{Advanced materials} \bibinfo{volume}{22} (\bibinfo{year}{2010}) \bibinfo{pages}{4249--4269}.
\bibitem[{Mirkin and Petrosko(2023)}]{mirkin2023inspired}
\bibinfo{author}{C.~A. Mirkin}, \bibinfo{author}{S.~H. Petrosko}, \bibinfo{title}{Inspired beyond nature: Three decades of spherical nucleic acids and colloidal crystal engineering with dna}, \bibinfo{year}{2023}.
\bibitem[{Ye et~al.(2015)Ye, Zhu, Ercius, Raja, He, Jones, Hauwiller, Liu, Xu, and Alivisatos}]{ye2015structural}
\bibinfo{author}{X.~Ye}, \bibinfo{author}{C.~Zhu}, \bibinfo{author}{P.~Ercius}, \bibinfo{author}{S.~N. Raja}, \bibinfo{author}{B.~He}, \bibinfo{author}{M.~R. Jones}, \bibinfo{author}{M.~R. Hauwiller}, \bibinfo{author}{Y.~Liu}, \bibinfo{author}{T.~Xu}, \bibinfo{author}{A.~P. Alivisatos},
\newblock \bibinfo{title}{Structural diversity in binary superlattices self-assembled from polymer-grafted nanocrystals},
\newblock \bibinfo{journal}{Nature communications} \bibinfo{volume}{6} (\bibinfo{year}{2015}) \bibinfo{pages}{10052}.
\bibitem[{Zhang et~al.(2013)Zhang, Lu, Yager, Van Der~Lelie, and Gang}]{zhang2013general}
\bibinfo{author}{Y.~Zhang}, \bibinfo{author}{F.~Lu}, \bibinfo{author}{K.~G. Yager}, \bibinfo{author}{D.~Van Der~Lelie}, \bibinfo{author}{O.~Gang},
\newblock \bibinfo{title}{A general strategy for the dna-mediated self-assembly of functional nanoparticles into heterogeneous systems},
\newblock \bibinfo{journal}{Nature nanotechnology} \bibinfo{volume}{8} (\bibinfo{year}{2013}) \bibinfo{pages}{865--872}.
\bibitem[{Hatton et~al.(2010)Hatton, Mishchenko, Davis, Sandhage, and Aizenberg}]{hatton2010assembly}
\bibinfo{author}{B.~Hatton}, \bibinfo{author}{L.~Mishchenko}, \bibinfo{author}{S.~Davis}, \bibinfo{author}{K.~H. Sandhage}, \bibinfo{author}{J.~Aizenberg},
\newblock \bibinfo{title}{Assembly of large-area, highly ordered, crack-free inverse opal films},
\newblock \bibinfo{journal}{Proceedings of the National Academy of Sciences} \bibinfo{volume}{107} (\bibinfo{year}{2010}) \bibinfo{pages}{10354--10359}.
\bibitem[{Marlow et~al.(2009)Marlow, Muldarisnur, Sharifi, Brinkmann, and Mendive}]{marlow2009opals}
\bibinfo{author}{F.~Marlow}, \bibinfo{author}{Muldarisnur}, \bibinfo{author}{P.~Sharifi}, \bibinfo{author}{R.~Brinkmann}, \bibinfo{author}{C.~Mendive},
\newblock \bibinfo{title}{Opals: status and prospects},
\newblock \bibinfo{journal}{Angewandte Chemie International Edition} \bibinfo{volume}{48} (\bibinfo{year}{2009}) \bibinfo{pages}{6212--6233}.
\bibitem[{Chen et~al.(2008)Chen, Wang, Wen, Zhang, Wang, Song, Jiang, and Zhu}]{chen2008fabrication}
\bibinfo{author}{X.~Chen}, \bibinfo{author}{L.~Wang}, \bibinfo{author}{Y.~Wen}, \bibinfo{author}{Y.~Zhang}, \bibinfo{author}{J.~Wang}, \bibinfo{author}{Y.~Song}, \bibinfo{author}{L.~Jiang}, \bibinfo{author}{D.~Zhu},
\newblock \bibinfo{title}{Fabrication of closed-cell polyimide inverse opal photonic crystals with excellent mechanical properties and thermal stability},
\newblock \bibinfo{journal}{Journal of Materials Chemistry} \bibinfo{volume}{18} (\bibinfo{year}{2008}) \bibinfo{pages}{2262--2267}.
\bibitem[{do~Ros{\'a}rio et~al.(2015)do~Ros{\'a}rio, Lilleodden, Waleczek, Kubrin, Petrov, Dyachenko, Sabisch, Nielsch, Huber, Eich et~al.}]{do2015self}
\bibinfo{author}{J.~J. do~Ros{\'a}rio}, \bibinfo{author}{E.~T. Lilleodden}, \bibinfo{author}{M.~Waleczek}, \bibinfo{author}{R.~Kubrin}, \bibinfo{author}{A.~Y. Petrov}, \bibinfo{author}{P.~N. Dyachenko}, \bibinfo{author}{J.~E. Sabisch}, \bibinfo{author}{K.~Nielsch}, \bibinfo{author}{N.~Huber}, \bibinfo{author}{M.~Eich}, et~al.,
\newblock \bibinfo{title}{Self-assembled ultra high strength, ultra stiff mechanical metamaterials based on inverse opals},
\newblock \bibinfo{journal}{Advanced Engineering Materials} \bibinfo{volume}{17} (\bibinfo{year}{2015}) \bibinfo{pages}{1420--1424}.
\bibitem[{Michelson et~al.(2023)Michelson, Flanagan, Lee, and Gang}]{michelson2023high}
\bibinfo{author}{A.~Michelson}, \bibinfo{author}{T.~J. Flanagan}, \bibinfo{author}{S.-W. Lee}, \bibinfo{author}{O.~Gang},
\newblock \bibinfo{title}{High-strength, lightweight nano-architected silica},
\newblock \bibinfo{journal}{Cell Reports Physical Science} \bibinfo{volume}{4} (\bibinfo{year}{2023}).
\bibitem[{Li et~al.(2022)Li, Zhou, Tanriover, Hadibrata, Partridge, Lin, Hu, Lee, Liu, Dravid et~al.}]{li2022open}
\bibinfo{author}{Y.~Li}, \bibinfo{author}{W.~Zhou}, \bibinfo{author}{I.~Tanriover}, \bibinfo{author}{W.~Hadibrata}, \bibinfo{author}{B.~E. Partridge}, \bibinfo{author}{H.~Lin}, \bibinfo{author}{X.~Hu}, \bibinfo{author}{B.~Lee}, \bibinfo{author}{J.~Liu}, \bibinfo{author}{V.~P. Dravid}, et~al.,
\newblock \bibinfo{title}{Open-channel metal particle superlattices},
\newblock \bibinfo{journal}{Nature} \bibinfo{volume}{611} (\bibinfo{year}{2022}) \bibinfo{pages}{695--701}.
\bibitem[{Lee et~al.(2022)Lee, Yee, Ye, and Macfarlane}]{lee2022nanoparticle}
\bibinfo{author}{M.~S. Lee}, \bibinfo{author}{D.~W. Yee}, \bibinfo{author}{M.~Ye}, \bibinfo{author}{R.~J. Macfarlane},
\newblock \bibinfo{title}{Nanoparticle assembly as a materials development tool},
\newblock \bibinfo{journal}{Journal of the American Chemical Society} \bibinfo{volume}{144} (\bibinfo{year}{2022}) \bibinfo{pages}{3330--3346}.
\bibitem[{Elghanian et~al.(1997)Elghanian, Storhoff, Mucic, Letsinger, and Mirkin}]{elghanian1997selective}
\bibinfo{author}{R.~Elghanian}, \bibinfo{author}{J.~J. Storhoff}, \bibinfo{author}{R.~C. Mucic}, \bibinfo{author}{R.~L. Letsinger}, \bibinfo{author}{C.~A. Mirkin},
\newblock \bibinfo{title}{Selective colorimetric detection of polynucleotides based on the distance-dependent optical properties of gold nanoparticles},
\newblock \bibinfo{journal}{Science} \bibinfo{volume}{277} (\bibinfo{year}{1997}) \bibinfo{pages}{1078--1081}.
\bibitem[{Lewis et~al.(2020)Lewis, Carter, and Macfarlane}]{lewis2020using}
\bibinfo{author}{D.~J. Lewis}, \bibinfo{author}{D.~J. Carter}, \bibinfo{author}{R.~J. Macfarlane},
\newblock \bibinfo{title}{Using dna to control the mechanical response of nanoparticle superlattices},
\newblock \bibinfo{journal}{Journal of the American Chemical Society} \bibinfo{volume}{142} (\bibinfo{year}{2020}) \bibinfo{pages}{19181--19188}.
\bibitem[{Mirkin et~al.(2020)Mirkin, Letsinger, Mucic, and Storhoff}]{mirkin2020dna}
\bibinfo{author}{C.~A. Mirkin}, \bibinfo{author}{R.~L. Letsinger}, \bibinfo{author}{R.~C. Mucic}, \bibinfo{author}{J.~J. Storhoff},
\newblock \bibinfo{title}{A dna-based method for rationally assembling nanoparticles into macroscopic materials},
\newblock in: \bibinfo{booktitle}{Spherical Nucleic Acids}, \bibinfo{publisher}{Jenny Stanford Publishing}, \bibinfo{year}{2020}, pp. \bibinfo{pages}{3--11}.
\bibitem[{Dey et~al.(2021)Dey, Fan, Gothelf, Li, Lin, Liu, Liu, Nijenhuis, Sacc{\`a}, Simmel et~al.}]{dey2021dna}
\bibinfo{author}{S.~Dey}, \bibinfo{author}{C.~Fan}, \bibinfo{author}{K.~V. Gothelf}, \bibinfo{author}{J.~Li}, \bibinfo{author}{C.~Lin}, \bibinfo{author}{L.~Liu}, \bibinfo{author}{N.~Liu}, \bibinfo{author}{M.~A. Nijenhuis}, \bibinfo{author}{B.~Sacc{\`a}}, \bibinfo{author}{F.~C. Simmel}, et~al.,
\newblock \bibinfo{title}{Dna origami},
\newblock \bibinfo{journal}{Nature Reviews Methods Primers} \bibinfo{volume}{1} (\bibinfo{year}{2021}) \bibinfo{pages}{13}.
\bibitem[{Castro et~al.(2011)Castro, Kilchherr, Kim, Shiao, Wauer, Wortmann, Bathe, and Dietz}]{castro2011primer}
\bibinfo{author}{C.~E. Castro}, \bibinfo{author}{F.~Kilchherr}, \bibinfo{author}{D.-N. Kim}, \bibinfo{author}{E.~L. Shiao}, \bibinfo{author}{T.~Wauer}, \bibinfo{author}{P.~Wortmann}, \bibinfo{author}{M.~Bathe}, \bibinfo{author}{H.~Dietz},
\newblock \bibinfo{title}{A primer to scaffolded dna origami},
\newblock \bibinfo{journal}{Nature methods} \bibinfo{volume}{8} (\bibinfo{year}{2011}) \bibinfo{pages}{221--229}.
\bibitem[{Hong et~al.(2017)Hong, Zhang, Liu, and Yan}]{hong2017dna}
\bibinfo{author}{F.~Hong}, \bibinfo{author}{F.~Zhang}, \bibinfo{author}{Y.~Liu}, \bibinfo{author}{H.~Yan},
\newblock \bibinfo{title}{Dna origami: scaffolds for creating higher order structures},
\newblock \bibinfo{journal}{Chemical reviews} \bibinfo{volume}{117} (\bibinfo{year}{2017}) \bibinfo{pages}{12584--12640}.
\bibitem[{Kulikowski et~al.(????)Kulikowski, Wang, Aitken, Wang, Doan, Lee, Zhang, Ke, and Gu}]{kulikowski4510528dna}
\bibinfo{author}{J.~Kulikowski}, \bibinfo{author}{S.~Wang}, \bibinfo{author}{Z.~H. Aitken}, \bibinfo{author}{M.~Wang}, \bibinfo{author}{D.~Doan}, \bibinfo{author}{A.~Lee}, \bibinfo{author}{Y.-W. Zhang}, \bibinfo{author}{Y.~Ke}, \bibinfo{author}{X.~W. Gu},
\newblock \bibinfo{title}{Dna-silica nanolattices as mechanical metamaterials},
\newblock \bibinfo{journal}{Available at SSRN 4510528}  (????).
\bibitem[{Peng et~al.(2022)Peng, Wu, Sun, Yue, Montgomery, Demoly, Zhou, Zhao, and Qi}]{peng20224d}
\bibinfo{author}{X.~Peng}, \bibinfo{author}{S.~Wu}, \bibinfo{author}{X.~Sun}, \bibinfo{author}{L.~Yue}, \bibinfo{author}{S.~M. Montgomery}, \bibinfo{author}{F.~Demoly}, \bibinfo{author}{K.~Zhou}, \bibinfo{author}{R.~R. Zhao}, \bibinfo{author}{H.~J. Qi},
\newblock \bibinfo{title}{4d printing of freestanding liquid crystal elastomers via hybrid additive manufacturing},
\newblock \bibinfo{journal}{Advanced Materials} \bibinfo{volume}{34} (\bibinfo{year}{2022}) \bibinfo{pages}{2204890}.
\bibitem[{Fu et~al.(2022)Fu, Li, Gong, Fan, Smith, Shen, Khalfalla, Huang, Qian, McCutcheon et~al.}]{fu20224d}
\bibinfo{author}{P.~Fu}, \bibinfo{author}{H.~Li}, \bibinfo{author}{J.~Gong}, \bibinfo{author}{Z.~Fan}, \bibinfo{author}{A.~T. Smith}, \bibinfo{author}{K.~Shen}, \bibinfo{author}{T.~O. Khalfalla}, \bibinfo{author}{H.~Huang}, \bibinfo{author}{X.~Qian}, \bibinfo{author}{J.~R. McCutcheon}, et~al.,
\newblock \bibinfo{title}{4d printing of polymers: Techniques, materials, and prospects},
\newblock \bibinfo{journal}{Progress in Polymer Science} \bibinfo{volume}{126} (\bibinfo{year}{2022}) \bibinfo{pages}{101506}.
\bibitem[{Sun et~al.(2024)Sun, Zhou, Demoly, Zhao, and Qi}]{sun2024perspective}
\bibinfo{author}{X.~Sun}, \bibinfo{author}{K.~Zhou}, \bibinfo{author}{F.~Demoly}, \bibinfo{author}{R.~R. Zhao}, \bibinfo{author}{H.~J. Qi},
\newblock \bibinfo{title}{Perspective: Machine learning in design for 3d/4d printing},
\newblock \bibinfo{journal}{Journal of Applied Mechanics} \bibinfo{volume}{91} (\bibinfo{year}{2024}).
\bibitem[{Jin et~al.(2021)Jin, Landauer, and Kim}]{jin2021ruga}
\bibinfo{author}{H.~Jin}, \bibinfo{author}{A.~K. Landauer}, \bibinfo{author}{K.-S. Kim},
\newblock \bibinfo{title}{Ruga mechanics of soft-orifice closure under external pressure},
\newblock \bibinfo{journal}{Proceedings of the Royal Society A} \bibinfo{volume}{477} (\bibinfo{year}{2021}) \bibinfo{pages}{20210238}.
\bibitem[{Leanza et~al.(2023)Leanza, Wu, Sun, Qi, and Zhao}]{leanza2023active}
\bibinfo{author}{S.~Leanza}, \bibinfo{author}{S.~Wu}, \bibinfo{author}{X.~Sun}, \bibinfo{author}{H.~J. Qi}, \bibinfo{author}{R.~R. Zhao},
\newblock \bibinfo{title}{Active materials for functional origami},
\newblock \bibinfo{journal}{Advanced Materials}  (\bibinfo{year}{2023}) \bibinfo{pages}{2302066}.
\bibitem[{Kim et~al.(2018)Kim, Yuk, Zhao, Chester, and Zhao}]{kim2018printing}
\bibinfo{author}{Y.~Kim}, \bibinfo{author}{H.~Yuk}, \bibinfo{author}{R.~Zhao}, \bibinfo{author}{S.~A. Chester}, \bibinfo{author}{X.~Zhao},
\newblock \bibinfo{title}{Printing ferromagnetic domains for untethered fast-transforming soft materials},
\newblock \bibinfo{journal}{Nature} \bibinfo{volume}{558} (\bibinfo{year}{2018}) \bibinfo{pages}{274--279}.
\bibitem[{Sola et~al.(2023)Sola, Trinchi, and Hill}]{sola2023self}
\bibinfo{author}{A.~Sola}, \bibinfo{author}{A.~Trinchi}, \bibinfo{author}{A.~J. Hill},
\newblock \bibinfo{title}{Self-assembly meets additive manufacturing: Bridging the gap between nanoscale arrangement of matter and macroscale fabrication},
\newblock \bibinfo{journal}{Smart Materials in Manufacturing} \bibinfo{volume}{1} (\bibinfo{year}{2023}) \bibinfo{pages}{100013}.
\bibitem[{J{\"a}kel et~al.(2023)J{\"a}kel, Heymann, and Simmel}]{jakel2023multiscale}
\bibinfo{author}{A.~C. J{\"a}kel}, \bibinfo{author}{M.~Heymann}, \bibinfo{author}{F.~C. Simmel},
\newblock \bibinfo{title}{Multiscale biofabrication: Integrating additive manufacturing with dna-programmable self-assembly},
\newblock \bibinfo{journal}{Advanced Biology} \bibinfo{volume}{7} (\bibinfo{year}{2023}) \bibinfo{pages}{2200195}.
\bibitem[{Ramachandramoorthy et~al.(2015)Ramachandramoorthy, Bernal, and Espinosa}]{ramachandramoorthy2015pushing}
\bibinfo{author}{R.~Ramachandramoorthy}, \bibinfo{author}{R.~Bernal}, \bibinfo{author}{H.~D. Espinosa},
\newblock \bibinfo{title}{Pushing the envelope of in situ transmission electron microscopy},
\newblock \bibinfo{journal}{ACS nano} \bibinfo{volume}{9} (\bibinfo{year}{2015}) \bibinfo{pages}{4675--4685}.
\bibitem[{Bhowmick et~al.(2019)Bhowmick, Espinosa, Jungjohann, Pardoen, and Pierron}]{bhowmick2019advanced}
\bibinfo{author}{S.~Bhowmick}, \bibinfo{author}{H.~Espinosa}, \bibinfo{author}{K.~Jungjohann}, \bibinfo{author}{T.~Pardoen}, \bibinfo{author}{O.~Pierron},
\newblock \bibinfo{title}{Advanced microelectromechanical systems-based nanomechanical testing: beyond stress and strain measurements},
\newblock \bibinfo{journal}{Mrs Bulletin} \bibinfo{volume}{44} (\bibinfo{year}{2019}) \bibinfo{pages}{487--493}.
\bibitem[{Shaikeea et~al.(2022)Shaikeea, Cui, O’Masta, Zheng, and Deshpande}]{shaikeea2022toughness}
\bibinfo{author}{A.~J.~D. Shaikeea}, \bibinfo{author}{H.~Cui}, \bibinfo{author}{M.~O’Masta}, \bibinfo{author}{X.~R. Zheng}, \bibinfo{author}{V.~S. Deshpande},
\newblock \bibinfo{title}{The toughness of mechanical metamaterials},
\newblock \bibinfo{journal}{Nature materials} \bibinfo{volume}{21} (\bibinfo{year}{2022}) \bibinfo{pages}{297--304}.
\bibitem[{Mukherjee et~al.(2022)Mukherjee, Patino, Pathak, Lemaitre, and Espinosa}]{mukherjee2022deep}
\bibinfo{author}{P.~Mukherjee}, \bibinfo{author}{C.~A. Patino}, \bibinfo{author}{N.~Pathak}, \bibinfo{author}{V.~Lemaitre}, \bibinfo{author}{H.~D. Espinosa},
\newblock \bibinfo{title}{Deep learning-assisted automated single cell electroporation platform for effective genetic manipulation of hard-to-transfect cells},
\newblock \bibinfo{journal}{Small} \bibinfo{volume}{18} (\bibinfo{year}{2022}) \bibinfo{pages}{2107795}.
\bibitem[{Patino et~al.(2022)Patino, Pathak, Mukherjee, Park, Bao, and Espinosa}]{patino2022multiplexed}
\bibinfo{author}{C.~A. Patino}, \bibinfo{author}{N.~Pathak}, \bibinfo{author}{P.~Mukherjee}, \bibinfo{author}{S.~H. Park}, \bibinfo{author}{G.~Bao}, \bibinfo{author}{H.~D. Espinosa},
\newblock \bibinfo{title}{Multiplexed high-throughput localized electroporation workflow with deep learning--based analysis for cell engineering},
\newblock \bibinfo{journal}{Science Advances} \bibinfo{volume}{8} (\bibinfo{year}{2022}) \bibinfo{pages}{eabn7637}.
\bibitem[{Samanta et~al.(2019)Samanta, Iscen, Laramy, Ebrahimi, Bujold, Schatz, and Mirkin}]{samanta2019multivalent}
\bibinfo{author}{D.~Samanta}, \bibinfo{author}{A.~Iscen}, \bibinfo{author}{C.~R. Laramy}, \bibinfo{author}{S.~B. Ebrahimi}, \bibinfo{author}{K.~E. Bujold}, \bibinfo{author}{G.~C. Schatz}, \bibinfo{author}{C.~A. Mirkin},
\newblock \bibinfo{title}{Multivalent cation-induced actuation of dna-mediated colloidal superlattices},
\newblock \bibinfo{journal}{Journal of the American Chemical Society} \bibinfo{volume}{141} (\bibinfo{year}{2019}) \bibinfo{pages}{19973--19977}.
\bibitem[{Hollingsworth and Dror(2018)}]{hollingsworth2018molecular}
\bibinfo{author}{S.~A. Hollingsworth}, \bibinfo{author}{R.~O. Dror},
\newblock \bibinfo{title}{Molecular dynamics simulation for all},
\newblock \bibinfo{journal}{Neuron} \bibinfo{volume}{99} (\bibinfo{year}{2018}) \bibinfo{pages}{1129--1143}.
\bibitem[{Jin et~al.(2023)Jin, Zhang, and Espinosa}]{jin2023recent}
\bibinfo{author}{H.~Jin}, \bibinfo{author}{E.~Zhang}, \bibinfo{author}{H.~D. Espinosa},
\newblock \bibinfo{title}{Recent advances and applications of machine learning in experimental solid mechanics: A review},
\newblock \bibinfo{journal}{arXiv preprint arXiv:2303.07647}  (\bibinfo{year}{2023}).
\bibitem[{Guo et~al.(2021)Guo, Yang, Yu, and Buehler}]{guo2021artificial}
\bibinfo{author}{K.~Guo}, \bibinfo{author}{Z.~Yang}, \bibinfo{author}{C.-H. Yu}, \bibinfo{author}{M.~J. Buehler},
\newblock \bibinfo{title}{Artificial intelligence and machine learning in design of mechanical materials},
\newblock \bibinfo{journal}{Materials Horizons} \bibinfo{volume}{8} (\bibinfo{year}{2021}) \bibinfo{pages}{1153--1172}.
\bibitem[{Alderete et~al.(2022)Alderete, Pathak, and Espinosa}]{alderete2022machine}
\bibinfo{author}{N.~A. Alderete}, \bibinfo{author}{N.~Pathak}, \bibinfo{author}{H.~D. Espinosa},
\newblock \bibinfo{title}{Machine learning assisted design of shape-programmable 3d kirigami metamaterials},
\newblock \bibinfo{journal}{npj Computational Materials} \bibinfo{volume}{8} (\bibinfo{year}{2022}) \bibinfo{pages}{191}.
\bibitem[{Schmidt and Lipson(2009)}]{schmidt2009distilling}
\bibinfo{author}{M.~Schmidt}, \bibinfo{author}{H.~Lipson},
\newblock \bibinfo{title}{Distilling free-form natural laws from experimental data},
\newblock \bibinfo{journal}{science} \bibinfo{volume}{324} (\bibinfo{year}{2009}) \bibinfo{pages}{81--85}.
\bibitem[{Mirjalili and Mirjalili(2019)}]{mirjalili2019genetic}
\bibinfo{author}{S.~Mirjalili}, \bibinfo{author}{S.~Mirjalili},
\newblock \bibinfo{title}{Genetic algorithm},
\newblock \bibinfo{journal}{Evolutionary Algorithms and Neural Networks: Theory and Applications}  (\bibinfo{year}{2019}) \bibinfo{pages}{43--55}.
\bibitem[{Karniadakis et~al.(2021)Karniadakis, Kevrekidis, Lu, Perdikaris, Wang, and Yang}]{karniadakis2021physics}
\bibinfo{author}{G.~E. Karniadakis}, \bibinfo{author}{I.~G. Kevrekidis}, \bibinfo{author}{L.~Lu}, \bibinfo{author}{P.~Perdikaris}, \bibinfo{author}{S.~Wang}, \bibinfo{author}{L.~Yang},
\newblock \bibinfo{title}{Physics-informed machine learning},
\newblock \bibinfo{journal}{Nature Reviews Physics} \bibinfo{volume}{3} (\bibinfo{year}{2021}) \bibinfo{pages}{422--440}.
\bibitem[{Raissi et~al.(2019)Raissi, Perdikaris, and Karniadakis}]{raissi2019physics}
\bibinfo{author}{M.~Raissi}, \bibinfo{author}{P.~Perdikaris}, \bibinfo{author}{G.~E. Karniadakis},
\newblock \bibinfo{title}{Physics-informed neural networks: A deep learning framework for solving forward and inverse problems involving nonlinear partial differential equations},
\newblock \bibinfo{journal}{Journal of Computational physics} \bibinfo{volume}{378} (\bibinfo{year}{2019}) \bibinfo{pages}{686--707}.
\bibitem[{Song and Jin(2023)}]{song2023identifying}
\bibinfo{author}{S.~Song}, \bibinfo{author}{H.~Jin},
\newblock \bibinfo{title}{Identifying constitutive parameters for complex hyperelastic solids using physics-informed neural networks},
\newblock \bibinfo{journal}{arXiv preprint arXiv:2308.15640}  (\bibinfo{year}{2023}).
\bibitem[{Zhang et~al.(2022)Zhang, Dao, Karniadakis, and Suresh}]{zhang2022analyses}
\bibinfo{author}{E.~Zhang}, \bibinfo{author}{M.~Dao}, \bibinfo{author}{G.~E. Karniadakis}, \bibinfo{author}{S.~Suresh},
\newblock \bibinfo{title}{Analyses of internal structures and defects in materials using physics-informed neural networks},
\newblock \bibinfo{journal}{Science advances} \bibinfo{volume}{8} (\bibinfo{year}{2022}) \bibinfo{pages}{eabk0644}.
\bibitem[{Lu et~al.(2021)Lu, Jin, Pang, Zhang, and Karniadakis}]{lu2021learning}
\bibinfo{author}{L.~Lu}, \bibinfo{author}{P.~Jin}, \bibinfo{author}{G.~Pang}, \bibinfo{author}{Z.~Zhang}, \bibinfo{author}{G.~E. Karniadakis},
\newblock \bibinfo{title}{Learning nonlinear operators via deeponet based on the universal approximation theorem of operators},
\newblock \bibinfo{journal}{Nature machine intelligence} \bibinfo{volume}{3} (\bibinfo{year}{2021}) \bibinfo{pages}{218--229}.
\bibitem[{Goswami et~al.(2023)Goswami, Bora, Yu, and Karniadakis}]{goswami2023physics}
\bibinfo{author}{S.~Goswami}, \bibinfo{author}{A.~Bora}, \bibinfo{author}{Y.~Yu}, \bibinfo{author}{G.~E. Karniadakis},
\newblock \bibinfo{title}{Physics-informed deep neural operator networks},
\newblock in: \bibinfo{booktitle}{Machine Learning in Modeling and Simulation: Methods and Applications}, \bibinfo{publisher}{Springer}, \bibinfo{year}{2023}, pp. \bibinfo{pages}{219--254}.
\bibitem[{Krenn et~al.(2022)Krenn, Pollice, Guo, Aldeghi, Cervera-Lierta, Friederich, dos Passos~Gomes, H{\"a}se, Jinich, Nigam et~al.}]{krenn2022scientific}
\bibinfo{author}{M.~Krenn}, \bibinfo{author}{R.~Pollice}, \bibinfo{author}{S.~Y. Guo}, \bibinfo{author}{M.~Aldeghi}, \bibinfo{author}{A.~Cervera-Lierta}, \bibinfo{author}{P.~Friederich}, \bibinfo{author}{G.~dos Passos~Gomes}, \bibinfo{author}{F.~H{\"a}se}, \bibinfo{author}{A.~Jinich}, \bibinfo{author}{A.~Nigam}, et~al.,
\newblock \bibinfo{title}{On scientific understanding with artificial intelligence},
\newblock \bibinfo{journal}{Nature Reviews Physics} \bibinfo{volume}{4} (\bibinfo{year}{2022}) \bibinfo{pages}{761--769}.

\end{thebibliography}

\end{document}